\documentclass[lettersize,journal]{IEEEtran}
\usepackage{amsmath,amsfonts}
\usepackage{algorithmic}
\usepackage{algorithm}
\usepackage{array}
\usepackage[caption=false,font=normalsize,labelfont=sf,textfont=sf]{subfig}
\usepackage{textcomp}
\usepackage{stfloats}
\usepackage{url}
\usepackage{verbatim}
\usepackage{graphicx}
\usepackage{cite}
\usepackage{hyperref}
\usepackage{soul}
\usepackage{xcolor}
\sethlcolor{yellow}
\begin{document}
\bstctlcite{IEEEexample:BSTcontrol}
\title{Hybrid Full Waveform Inversion Assisted by Rytov Approximation\\for Musculoskeletal Ultrasound Computed Tomography}

\author{
Yifei Sun, 
Yubing Li,~\IEEEmembership{Member,~IEEE}, 
Chang Su, 
Lekang Jiang, 
Xiangwei Lu, 
Ligang Cui,
He Sun,
and Weijun Lin%
\thanks{
This work is supported by the National Natural Science Foundation of China (No.~12474461), 
and by the Basic and Frontier Exploration Project Independently Deployed by the Institute of Acoustics, 
Chinese Academy of Sciences (No.~JCQY202402).
}
\thanks{
This work involved human subjects in its research. Approval of all ethical and experimental procedures and protocols was granted by the Institutional Review Board (IRB) of the Peking University Third Hospital under Approval No. IRB00006761-M2024690.
}
\thanks{
Corresponding authors: Yubing Li (e-mail: liyubing@mail.ioa.ac.cn), He Sun (email: hesun@pku.edu.cn), Weijun Lin (e-mail: linwj@mail.ioa.ac.cn).
}
\thanks{
Yifei Sun, Yubing Li, Chang Su, Lekang Jiang, Xiangwei Lu and Weijun Lin are with the Institute of Acoustics, Chinese Academy of Sciences, Beijing 100190, China, 
and the University of Chinese Academy of Sciences, Beijing 100049, China. Ligang Cui is with the Department of Ultrasound, Peking University Third Hospital, Beijing 100191, China. He Sun is with the College of Future Technology, Peking University, Beijing 100871, China, and the National Biomedical Imaging Center, Peking University, Beijing 100871, China.
}
\thanks{
Yifei Sun is also with Université Bourgogne Europe, IMVIA UR 7535, 21000 Dijon, France.
}
}



\maketitle

\begin{abstract}
Ultrasound computed tomography is emerging as a promising safe and accessible modality for soft-tissue medical imaging, with full waveform inversion playing a key role in unlocking its full potential for high-resolution, quantitative reconstructions. Frequency domain full waveform inversion (FDFWI) for reconstructing spatial maps of acoustic properties in the musculoskeletal system is highly sensitive to the quality of low-frequency signals, making the final imaging outcome vulnerable to issues such as inappropriate initial models and strong scatterings related to bones. To address these challenges, we propose a hybrid full waveform inversion (HFWI) algorithm that incorporates a traveltime inversion algorithm based on the generalized Rytov approximation into the FDFWI framework. This hybrid strategy enhances early-stage inversion quality and substantially reduces sensitivity to the initial model, all while maintaining computational efficiency. Importantly, HFWI achieves results comparable to those obtained using well-constructed initial models, without incurring extra computational cost, thus enabling accurate imaging under realistic, bandwidth-limited conditions. In addition, we introduce a near real-time strategy to update first-arrival traveltimes based on forward-scattered phase variations without requiring extra wavefield simulations. Numerical simulations, as well as \textit{in vitro} and \textit{in vivo} experiments  confirm the robustness and efficiency of the proposed approach. HFWI also shows promise to extend to more complex scenarios of musculoskeletal parametric reconstruction. 
\end{abstract}

\begin{IEEEkeywords}
Musculoskeletal system, ultrasound computed tomography, full waveform inversion, quantitative ultrasound.
\end{IEEEkeywords}

\section{Introduction}
\IEEEPARstart{I}{n} sports medicine, assessing the musculoskeletal system relies predominantly on medical imaging modalities, with ultrasound computed tomography (USCT) recently emerging as a promising technique \cite{ref2,ref3,ref5,ref7}. USCT is a quantitative ultrasound (QUS) technique \cite{ref9,ref10} that reconstructs acoustic parameters such as sound speed, density, and attenuation \cite{ref11}. These parameters have been shown to correlate with physiological changes in soft tissue \cite{ref12,ref13}, whose diagnosis remains challenging in sports medicine. Compared with X-ray computed tomography (X-CT) and magnetic resonance imaging (MRI), USCT is radiation-free, metal-insensitive, and low-cost—making it highly attractive in sports medicine. Recent researches indicate that USCT can achieve high-resolution imaging of soft tissues \cite{ref18,ref19}, and has been extensively studied in breast cancer detection \cite{ref11,ref21,ref23}. Nonetheless, extending USCT to the musculoskeletal system remains challenging due to the high acoustic impedance and complex structure of bone tissues. Although certain ultrasound-based bone studies have been reported \cite{ref2,ref3,ref7}, a clinically ready USCT framework that fully exploits its strength in soft-tissue assessment within the broader musculoskeletal context is still under development.

In USCT, acoustic parameter distributions (i.e., images) are reconstructed from time-series data collected by transducer arrays. USCT algorithms are broadly classified into two categories. The first focuses on travel-time information, exemplified by time-of-flight tomography (TOFT) \cite{ref21,ref24} and reflection tomography \cite{ref25,ref26}. These methods estimate the sound speed model by minimizing discrepancies between observed and synthetic traveltimes, computed via the Eikonal equation. The second formulates and resolves the inverse problem using the waveform information of recorded signals. For instance, diffraction tomography (DT) \cite{ref18} addresses diffraction and scattering inverse problems via linearized wavefield approximations, while full-waveform inversion (FWI) \cite{ref27,ref28,ref29} employs the full waveform, including higher-order effects, for model reconstruction. Recent studies suggest FWI is promising for bone imaging \cite{ref30,ref31}, given sufficiently dense data. This typically requires a ring or cylindrical array with full-matrix capture (FMC) acquisition.

FWI has been extensively studied in seismic imaging, medical imaging \cite{ref32,ref34}, and nondestructive testing \cite{ref37}. It defines its loss function as the mismatch between observed and synthetic waveforms for each transmitter–receiver pair, theoretically achieving resolution down to half the wavelength \cite{ref38}. FWI can be categorized into time-domain (TDFWI) \cite{ref39} and frequency-domain (FDFWI) \cite{ref40,ref41} approaches, mainly differing in how wavefields are simulated. In both seismology and medical imaging, most studies have focused on TDFWI \cite{ref34,ref44}. However, TDFWI relies on time-domain solvers \cite{ref46,ref48} that must satisfy the Courant condition \cite{ref49} for numerical stability. This imposes a heavy computational load, especially when the model spans hundreds of wavelengths, requiring significant resources for reasonable runtimes. While such demands are acceptable in geophysics, they remain a major bottleneck for USCT, where clinical feasibility requires faster, more efficient imaging. Consequently, the practical use of TDFWI in USCT is still hindered by its computational cost.

By comparison, frequency-domain methods offer a promising alternative when focusing on a finite number of discrete frequency points, since doing so curtails the amount of information required to solve the wave equation \cite{ref50,ref51}. FDFWI has been employed in breast imaging \cite{ref52} and its application in musculoskeletal imaging remains largely untapped. By restricting the inversion to a finite set of discrete frequencies in practice, FDFWI reduces the nonlinearity of the inversion compared to TDFWI, since each monochromatic inversion is better conditioned. However, unless an appropriate frequency continuation strategy is employed, the lack of explicit broadband phase consistency can still render the objective function highly nonconvex and prone to cycle skipping \cite{ref53,ref54}.This typically arises when the phase mismatch between observed and synthetic waveforms exceeds half a cycle—a situation often exacerbated by the high sound speed in bone, especially when starting from a homogeneous initial model.

Two main approaches have been proposed to mitigate cycle skipping in FDFWI. One exploits low-frequency data \cite{ref54,ref55}, which increase the effective wavelength and motivate multiscale strategies in which low-frequency reconstructions are used to initialize higher-frequency inversions \cite{ref30,ref57,ref_revision_2}; however, the availability and quality of low-frequency signals are often limited by hardware constraints. The other approach focuses on constructing improved initial models \cite{ref56}, typically through sequential hybrid workflows that use linearized methods based on reflected waves \cite{ref57} or traveltime information \cite{ref59}. In media with strong sound-speed and impedance contrasts, such as musculoskeletal tissues, these methods may require careful parameter tuning to ensure stable inversion. Alternatively, a simultaneous hybrid strategy has been recently introduced, in which traveltime-difference terms are incorporated directly into the waveform misfit \cite{ref_revision_1}, at the expense of additional computational cost due to repeated Eikonal solves.

To reduce the computational overhead associated with both the initialization stage of sequential hybrid approaches and the Eikonal-based simultaneous hybrid methods, while improving reconstruction quality, we propose to incorporate a Rytov-based formulation—commonly used in optics—into the simultaneous hybrid inversion framework. By introducing an exponential multiplicative factor, the Rytov approximation creates a linear relationship between the complex wave phase and the model perturbation \cite{ref60}, allowing it to accommodate large phase shifts more effectively than the traditional linearization. Building on this foundation, Feng et al. (2019) further advanced this concept by introducing a generalized Rytov approximation (GRA) \cite{ref61}, which accurately models the phase behavior of forward-scattered waves even under strong sound speed perturbations. Leveraging GRA, inversion methods \cite{ref62,ref63} have incorporated traveltime sensitivity kernels \cite{ref65} to reconstruct preliminary models that serves as effective initial inputs for FWI. These works provide a foundation for designing a simultaneous hybrid framework tailored to USCT-FWI in the musculoskeletal system, where robust initialization remains a key challenge.

In this work, we propose to integrate the Generalized Rytov Approximation–based sensitivity kernel (GRA-TSK) \cite{ref62,ref63}, originally developed for time-domain, band-limited formulations, into FDFWI operating at discrete frequency points. This integration leads to a simultaneous hybrid inversion with a unified misfit function that combines waveform-disperency and traveltime-error terms, enabling robust early-stage updates. The proposed method differs from conventional FDFWI only in the misfit definition and the associated adjoint source, which incorporates Rytov-based kinematic information, while the forward and adjoint solvers remain unchanged. Consequently, no additional computational overhead is introduced relative to standard FDFWI.

The paper is organized as follows. First, we introduce the scanning hardware, and detail the formulations of both GRA-TSK and FDFWI, followed by the proposal of a unified strategy. The new optimization scheme simultaneously performs Rytov and FWI updates at each inversion step, and the hybrid algorithm only incurs negligible additional computation costs compared to standard FDFWI. Next, we validate the proposed method through numerical simulations and \textit{in vitro} experiments, demonstrating reduced dependency on initial models and its improved handling of complex musculoskeletal structures. Finally, our \textit{in vivo} explorations further illustrate the practical efficiency gains offered by the proposed method.

\section{Methods}
\subsection{Scanning Hardware}

The experimental data are acquired using a 512-element ring array integrated with a self-developed 256-channel ultrasound acquisition system. The ring array consists of four 128-element subarrays, with all elements evenly distributed along a circular aperture of 22 cm in diameter (Fig.~1(a)). Each transducer has a center frequency of approximately 0.9 MHz and operates in both transmit and receive modes, enabling full-matrix capture (FMC) over all transmitter--receiver pairs.

The acquisition system contains eight 32-channel boards, providing 256 channels in total. Each board uses a time-division multiplexing (TDM) strategy to control up to 128 transducers through programmable switching. In this study, four boards are connected to the ring array, and FMC over all 512 elements is realized through 2048 transmit events. The time-series data are sampled at 25\,MHz.

To obtain usable low-frequency signals, the array is excited by a single-cycle bipolar square wave with a fundamental frequency of 0.6 MHz, yielding an effective bandwidth of 0.25--1.2\,MHz, including the low-frequency components required for robust inversion. Each transducing element has a height of approximately 2 cm, providing strong vertical directivity and concentrating acoustic energy within the imaging plane. The ring array is mounted on a mechanical lift and fully immersed in water (Fig.~1(b)), with the object placed near the geometric center. Different transverse slices are acquired by adjusting the vertical position of the lift.


\begin{figure}[H]
\centering
\vspace{-3mm}

\includegraphics[width=3.5in]{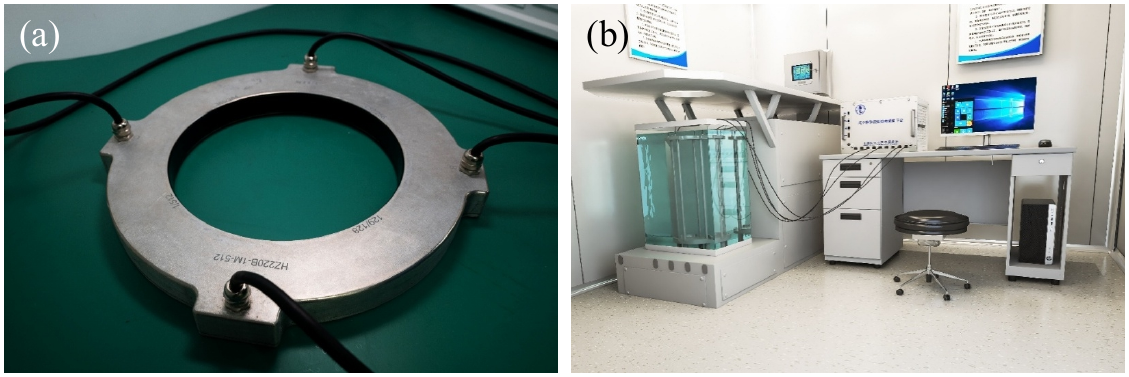}

\vspace{-2mm}
\caption{Experimental setup of USCT. (a) Ring array transducer. (b) Rendering of imaging system with water tank, motion platform, and acquisition unit.}
\vspace{-2mm}
\label{fig_sim}

\end{figure}

\subsection{Inversion Technique}
The proposed inversion technique combines FDFWI with GRA-TSK-based traveltime inversion. GRA-TSK's robustness is used to guide FDFWI during early image reconstruction, improving initial model building. This section outlines the principles of both algorithms, introduces a traveltime update strategy for GRA-TSK within the FDFWI framework, and describes the hybrid FWI (HFWI) implementation.

\subsubsection{Full Waveform Inversion}
FWI iteratively updates model parameters by minimizing a misfit function that quantifies the difference between observed and synthetic waveforms. In conventional FDFWI, this is typically measured by the $L_2$-norm of the difference at a given frequency. Assuming an isotropic acoustic medium with constant density, the misfit function at angular frequency $\omega_f$ is:
\begin{equation}
J_{\mathrm{FWI}}(\mathbf{m}) = \frac{1}{2} \sum_s \sum_r \left( \sum_{\mathbf{x}} \delta(\mathbf{x}, \mathbf{r}) u_{\mathrm{syn}}(\mathbf{x}, \omega_f; \mathbf{s})-d(\mathbf{r}, \omega_f; \mathbf{s}) \right)^2,
\label{eq:fwi_cost}
\end{equation}
subject to the wave equation
\begin{equation}
\label{eq:wave_equation}
L(\omega_f, \mathbf{m}) u_{\mathrm{syn}}(\mathbf{x}, \omega_f; \mathbf{s}) = -f(\omega_f) \delta(\mathbf{x} - \mathbf{s}),
\end{equation}
where $L(\omega, \mathbf{m}) = \nabla^2 + \omega^2 \mathbf{m}^2(\mathbf{x})$ is the forward operator, $u_{\mathrm{syn}}$ is the synthetic wavefield, $d$ the observed data, $\mathbf{m}$ the slowness, and $f$ the source intensity. $\mathbf{x}$, $\mathbf{r}$, and $\mathbf{s}$ denote the imaging point, receiver, and source positions, respectively. $\nabla^2$ is the Laplacian operator and $\delta(\cdot)$ the Dirac delta function.

Using the adjoint method \cite{ref40}, we can obtain the gradient of $J_{FWI}$ with respect to the model parameter $\mathbf{m}$ as follows:
\begin{equation}
\nabla_{\mathbf{m}} J_{\mathrm{FWI}}(\mathbf{x}) = \operatorname{Re} \left( 
\sum_s \sum_r 2\,\omega^2\mathbf{m} \lambda_s^*(\mathbf{x}, \omega_f; \mathbf{s}) \, 
u_{\mathrm{syn}}(\mathbf{x}, \omega_f; \mathbf{s}) \right)
\label{eq:fwi_grad}
\end{equation}
with
\begin{equation}
L(\omega_f, \mathbf{m}) \lambda_s(\mathbf{x}, \omega_f; \mathbf{s}) 
= -f(\omega_f)\, \delta d(\mathbf{r}, \omega_f; \mathbf{s}),
\label{eq:backword_equation}
\end{equation}
where $\lambda_s$ is the adjoint wavefield backward propagated from the receiving points by setting the data residual $\delta d(\mathbf{r}, \omega_f; \mathbf{s}) = 
\sum_{\mathbf{x}} \delta(\mathbf{x}, \mathbf{r}) u_{\mathrm{syn}}(\mathbf{x}, \omega_f; \mathbf{s}) - d(\mathbf{r}, \omega_f; \mathbf{s})$ as the dummy sources. The gradient of the misfit function is obtained by the conjugate product of the forward-propagation field (originating from the sound source) and the backward-propagation field (emanating from the receiver). These two wavefields can be obtained by solving the wave equation with a numerical solver (our choice is introduced later).

By employing the gradient expression in~\eqref{eq:fwi_grad}, the minimization problem is solved through iterative local optimization:
\begin{equation}
\mathbf{m}_{k+1} = \mathbf{m}_k - \alpha_k \nabla_{\mathbf{m}} J_{\mathrm{FWI}},
\label{eq:update}
\end{equation}
where $k$ is the iteration index, $\alpha_k$ the step size determined by line search satisfying the Wolfe conditions \cite{ref67} and the initial model $\mathbf{m}_0$ (slowness) is typically set to homogeneous water.

To mitigate the inherent non-linearity---particularly the risk of cycle skipping---we adopt a multi-scale strategy \cite{ref_revision_2}. The inversion runs on a limited set of discrete frequency points; for each point, iterations continue until the misfit drops below a threshold or reaches the maximum count, after which the algorithm advances to the next (higher) frequency.

Although this schedule alleviates cycle skipping, the musculoskeletal setting remains vulnerable because practical systems seldom record sufficiently low-frequency data with adequate SNR. We therefore embed the GRA-TSK traveltime update into the early stages to stabilise FDFWI under such conditions.

\subsubsection{Generalized Rytov Approximation-Traveltime Inversion}
This section details the GRA-TSK traveltime inversion. Unlike FWI, it updates the model by minimizing first-arrival traveltime residuals between observed and synthetic data. We first derive the corresponding sensitivity kernel.

In scattering theory, the total field (as mentioned in~\eqref{eq:wave_equation}) is often decomposed into a background component and a scattered component. The background field is typically the pressure field associated with a reference slowness model $\mathbf{m}_{\mathrm{ref}}$. Formally, the background field is then expressed as:
\begin{equation}
L(\omega, \mathbf{m}_{\mathrm{ref}}) G_0(\mathbf{x}, \omega; \mathbf{s}) = -\delta(\mathbf{x} - \mathbf{s})
\label{eq:background_field}
\end{equation}
where $G_0(\mathbf{x}, \omega; \mathbf{s})$ represents the background Green’s function. Unlike the Born approximation, which linearizes the wave equation by assuming the scattered field is small and the total field is approximated as the sum of the incident and scattered fields, the Rytov approximation models the total field as a multiplicative perturbation of the incident field\cite{ref60}:
\begin{equation}
u(\mathbf{x}, \omega) = u_0(\mathbf{x}, \omega) e^{\psi(\mathbf{x}, \omega)}
\label{eq:incident_field}
\end{equation}
Here, $u_0$  is the incident field, and $\psi$ captures the complex phase perturbation. The Rytov approximation is valid when the phase variations induced by scattering remain moderate, which is typically governed by the condition\cite{ref60}
\begin{equation}
|\nabla \psi \cdot \nabla \psi| \ll \omega^2 |\Delta \mathbf{m}(\mathbf{x})^2|.
\label{eq:rytov_condition}
\end{equation}
where $\Delta \mathbf{m}(\mathbf{x}) = m_{\mathrm{true}}(\mathbf{x}) - \mathbf{m}_{\mathrm{ref}}(\mathbf{x})$ is the perturbation in slowness, and $m_{\mathrm{true}}(\mathbf{x})$ represents the true slowness distribution of the target.

However, this assumption is less suitable for acoustic fields that include bone tissue. By restricting scattering angles to the forward-scattering region, Feng et al. (2019) \cite{ref61} proposed GRA that linearizes slowness in terms of the first-arrival traveltime, thereby relaxing the constraint of weak slowness perturbations. Under GRA, the complex wave phase can be approximated by
\begin{equation}
\psi^{\mathrm{GRA}}(\mathbf{x}, \omega) =
\int \Delta \mathbf{m}(\mathbf{x}') \cdot
\frac{ 2\omega^2 \mathbf{m}_{\mathrm{ref}}(\mathbf{x}')
G_0(\mathbf{x}', \omega; \mathbf{x}) u_0(\mathbf{x}', \omega) }
{ u_0(\mathbf{x}, \omega) } \, \mathrm{d}\mathbf{x}'.
\label{eq:psi_gra}
\end{equation}
Feng et al. (2020) \cite{ref62} combine the complex phase in GRA with the traveltime inverse problem, formulating it as a least-squares misfit function:
\begin{equation}
J_{\mathrm{GRA}}(\mathbf{m}) = \sum_s \sum_r \Delta t^2,
\label{eq:gra_cost}
\end{equation}
where $\Delta t$ represents the difference between the first-arrival traveltimes in observed data and synthetic data. According to~\eqref{eq:incident_field}, we have
\begin{equation}
\Delta t = -\frac{\operatorname{Im} \psi(\mathbf{x}, \omega; \mathbf{s})}{\omega}.
\label{eq:delta_t}
\end{equation}
Then the $\Delta t$ can be expressed as
\begin{equation}
\Delta t(\mathbf{r}, \mathbf{s}) = \int \Delta \mathbf{m}(\mathbf{x})\, \mathbf{K}^{\mathrm{GRA}}(\mathbf{x}; \mathbf{r}, \mathbf{s}) \, \mathrm{d}\mathbf{x},
\label{eq:dt_kernel}
\end{equation}
where GRA-TSK is defined as
\begin{equation}
\mathbf{K}^{\mathrm{GRA}}(\mathbf{x}; \mathbf{r}, \mathbf{s}) = \int \mathrm{Re} \left[ 
- \frac{2 \omega \, \mathbf{m}_{\mathrm{ref}}(\mathbf{x}) \, G_0(\mathbf{x}, \omega; \mathbf{r}) \, u_0(\mathbf{x}, \omega; \mathbf{s})}
{i\,u_0(\mathbf{r}, \omega; \mathbf{s})}
\right] \mathrm{d}\omega.
\label{eq:K_GRA}
\end{equation}

Following the single-frequency formulation of FDFWI, where $m_{\mathrm{ref}}$, $u_0$, and $G_0$ denote the slowness and the associated wavefields corresponding to the current model in the inversion, the gradient of the misfit function~\eqref{eq:gra_cost} with respect to $\mathbf{m}$ can be written as:

\begin{equation}
\begin{aligned}
\nabla_{\mathbf{m}} J_{\mathrm{GRA}}(\mathbf{x}, \omega) &= \sum_{s} \sum_{r} \mathbf{K}^{\mathrm{GRA}}(\mathbf{x}, \omega; \mathbf{r}, \mathbf{s}) \, \Delta t(\mathbf{r}, \mathbf{s}) \\
&= \mathrm{Re} \left( \sum_{s} \sum_{r} \omega \mu_{\mathrm{GRA}}(\mathbf{x}, \omega; \mathbf{s}) \, u_{\mathrm{syn}}(\mathbf{x}, \omega; \mathbf{s}) \right),
\end{aligned}
\label{eq:grad_J_GRA}
\end{equation}
subject to
\begin{equation}
\mu_{\mathrm{GRA}} = -\frac{2 \, \mathbf{m}(\mathbf{x}) \, \Delta t(\mathbf{r}, \mathbf{s})}{i \, u_{\mathrm{syn}}(\mathbf{r}, \omega; \mathbf{s})} \, G_{\mathrm{syn}}(\mathbf{x}, \omega; \mathbf{r}).
\label{eq:mu_GRA}
\end{equation}

Similar to equation~\eqref{eq:update}, the minimization of misfit function~\eqref{eq:gra_cost} is accomplished through the following iterative local optimization scheme:
\begin{equation}
\mathbf{m}_{k+1} = \mathbf{m}_k - \alpha_k \nabla_{\mathbf{m}} J_{\mathrm{GRA}},
\label{eq:update_GRA}
\end{equation}
where the inversion at each frequency continues until the misfit satisfies the stopping criteria, after which the algorithm proceeds to the next frequency. 

Although GRA-TSK based traveltime inversion also incorporates certain waveform information, it primarily relies on kinematic information under a linearized approximation of phases for the transmitted signals, which generally enhances the robustness of the related inverse problem\cite{ref63}. 


\subsubsection{Hybrid Full Waveform Inversion}

In practice, robust FWI is challenging without sufficiently low-frequency signals or a reliable initial model. In contrast, GRA-TSK offers accurate early-stage reconstructions. Moreover, it shares the same forward solver (here, convergent Born series, CBS \cite{ref68}) and optimization strategy (here, a nonlinear conjugate-gradient (NCG) method; results obtained with an alternative optimizer are compared in Supplementary Material Section II) with FDFWI, enabling seamless integration into a hybrid framework, namely HFWI.

HFWI is formulated by combining the misfit functions of FWI and GRA-TSK. The model is updated by computing the gradient of the hybrid misfit function $J_{\mathrm{HFWI}}$ with respect to the model parameters $\mathbf{m}(\mathbf{x})$. We define the hybrid misfit function as

HFWI combines the misfit functions of FWI and GRA-TSK. The hybrid objective function is defined as:
\begin{equation}
J_{\mathrm{HFWI}} = (1 - \alpha) J_{\mathrm{FWI}} + \alpha \sigma J_{\mathrm{GRA}},
\label{eq:HFWI_cost}
\end{equation}
where $\alpha \in [0,1]$ is a hyperparameter controlling the relative contribution of each method to the model update, thereby balancing robustness and accuracy. The scaling factor $\sigma$ ensures that the two misfit values are on a comparable magnitude scale. The corresponding gradient of HFWI reads
\begin{equation}
G_{\mathrm{HFWI}} = (1 - \alpha) \nabla_{\mathbf{m}} J_{\mathrm{FWI}} + \alpha \sigma \nabla_{\mathbf{m}} J_{\mathrm{GRA}}.
\label{eq:HFWI_grad}
\end{equation}
Because the physical quantities in these two misfit functions differ significantly, a direct adjustment to unify them is challenging. Instead, we numerically align their magnitudes by setting:
\begin{equation}
\sigma = \frac{\lVert \nabla_{\mathbf{m}} J_{\mathrm{FWI}} \rVert}{\lVert \nabla_{\mathbf{m}} J_{\mathrm{GRA}} \rVert},
\label{eq:sigma}
\end{equation}
Notably, the adjoint sources used to compute gradients in both FWI and GRA-TSK share structural similarities. Combining equations~\eqref{eq:fwi_grad} and \eqref{eq:grad_J_GRA}, the gradient of HFWI with respect to slowness can be written as
\begin{equation}
\nabla_{\mathbf{m}} J_{\mathrm{HFWI}}(\mathbf{m}, \omega) = \sum_{s} \sum_{r} \mathrm{Re} \left( U_{\mathrm{adj}}(\mathbf{x}, \omega; \mathbf{s}) u_{\mathrm{syn}}(\mathbf{x}, \omega; \mathbf{s}) \right),
\label{eq:HFWI_grad_detail}
\end{equation}
where
\begin{equation}
U_{\mathrm{adj}} = (1 - \alpha) 2\, \omega^2 \mathbf{m}\lambda_s^*(\mathbf{x}, \omega; \mathbf{s}) + \alpha \sigma \omega \mu_{\mathrm{GRA}}.
\label{eq:HFWI_grad_adj}
\end{equation}
Equations~\eqref{eq:HFWI_grad_detail} and~\eqref{eq:HFWI_grad_adj} show that HFWI seamlessly integrates the forward modeling and optimization procedures of both methods. Wavefield computation (highlighted in the workflow of Fig. 2 as forward solver) is the primary numerical bottleneck in both FWI and GRA-TSK. The hybrid approach is designed to share these computed wavefields between the two algorithms. This ensures that the overall computational cost of HFWI remains comparable to that of each individual method. The workflow of HFWI is illustrated in Fig.2. By leveraging the convex optimization characteristics of GRA-TSK based traveltime inversion, this hybrid strategy helps mitigate the risk of FWI becoming trapped in local minima.

\begin{figure}[H]
\centering
\includegraphics[width=0.85\linewidth]{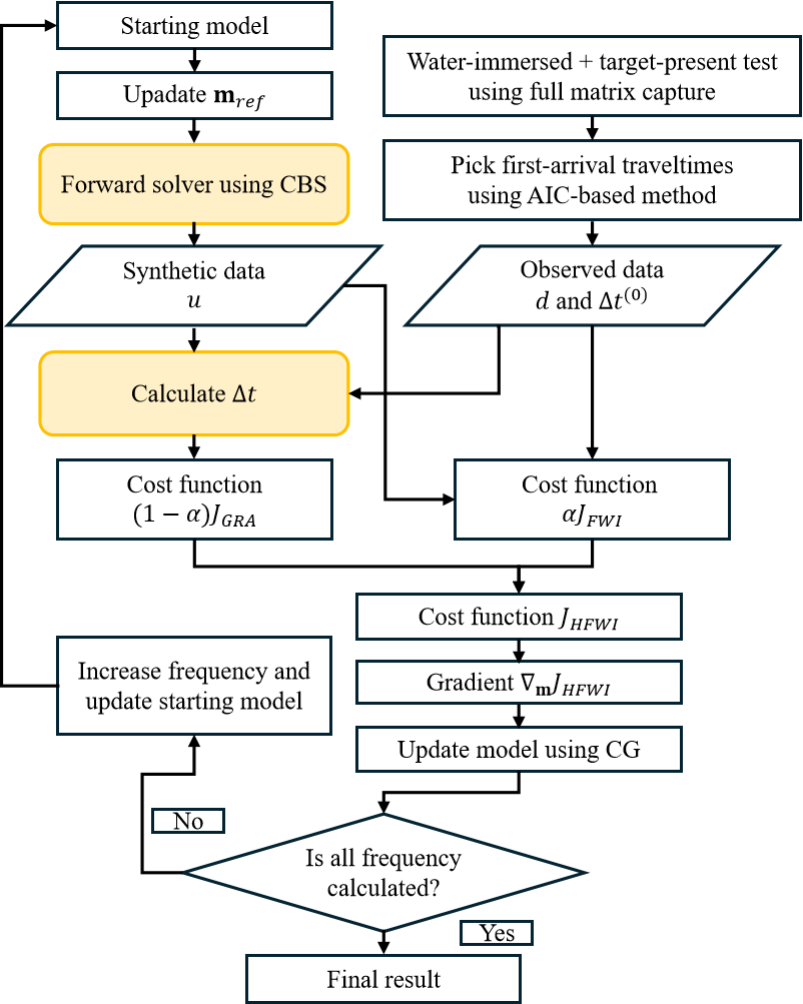}
\label{fig_2}
\vspace{-2mm}
\caption{Workflow of the proposed HFWI algorithm.}
\end{figure}

\subsubsection{Optimization of Traveltime Difference Extraction}

Beyond wavefield simulation, updating the traveltime difference $\Delta t$ (Fig. 2) is the other major computational bottleneck in HFWI. Classical GRA-based traveltime inversion derives first-arrival $\Delta t$ by matching time-domain synthetic and recorded traces \cite{ref62,ref63}. Because FDFWI generates no time-domain fields, this route is unavailable. Computing traveltimes from an Eikonal solver is possible, but adds considerable complexity and cost. To avoid these burdens, we develop a near-real-time algorithm that extracts $\Delta t$ directly within the frequency-domain loop, retaining accuracy while eliminating the overhead.


Equation \eqref{eq:delta_t} exploits the GRA linearisation of forward scattering to compute $\Delta t$ efficiently in the frequency domain. Because the complex-phase argument is periodic, the estimate is restricted to $\Delta t$ in $[-T/2, T/2]$, with $T=2\pi/\omega$ the period at the current frequency. In musculoskeletal imaging these shifts often lie outside that window, so using Eq. \eqref{eq:delta_t} unmodified can misestimate traveltimes.

To resolve this limitation, we introduce a two-step, near-real-time update of the traveltime map $\Delta t$, exploiting phase evolution between successive iterations (superscript $k$ denotes the current model). 
First, a short- and long-time average ratio (STA/LTA) picker \cite{ref69} extracts first-arrival times from the raw FMC data and forms the initial map $\Delta t^{(0)}$ with respect to a homogeneous-water reference. 
Second, as indicated by “Calculate $\Delta t$’’ in Fig. 2, $\Delta t^{(k)}$ is refined using the phase-derived correction between the synthetic responses at iterations $k$ and $k+1$. 
The update rule is

\begin{equation}
\Delta t_{N}^{(k+1)} = \Delta t_{N}^{(k)} + \mathrm{d}\big(\Delta t_{N}^{(k+1)}\big) + nT,
\label{eq:dt_update}
\end{equation}
where $N=1,\dots,512$ indexes the transducers. The updated traveltime difference map $\Delta t_{N}^{(k)}$ comprises two terms. The first, represented by the second term on the right-hand side of equation~\eqref{eq:dt_update}, is a phase-derived correction obtained from the change in synthetic wavefields between successive iterations:
\begin{equation}
\mathrm{d}\left( \Delta t_{N}^{(k+1)} \right) = -\frac{\mathrm{Im} \left( \ln \left( \frac{u_{N}^{(k+1)}(\omega; \mathbf{s})}{u_{N}^{(k)}(\omega; \mathbf{s})} \right) \right)}{\omega},
\label{eq:dt_increment}
\end{equation}
where $u_N^{(k)}$ is the response of transducer $N$ at iteration $k$. The second component, $nT$ in~\eqref{eq:dt_update}, unwraps phase by adding an integer multiple of the period $T$. For each transduce, an integer $n$ is chosen to keep phase variation smooth across neighbouring receivers:
\begin{equation}
\min_{n} \left\{ \left| nT + \mathrm{d}\left( \Delta t_{N}^{(k+1)} \right) - \mathrm{d}\left( \Delta t_{N-1}^{(k+1)} \right) \right| \right\}.
\label{eq:optimize_n}
\end{equation}
This correction assumes spatial smoothness in phase differences across neighbouring elements: a slowness update can be viewed as introducing a small perturbation into the previous reference model, and the receiver directly facing the source may experience a phase shift exceeding one period, yet the shift difference between adjacent receivers stays small thanks to the dense ring array. This continuity guarantees a coherent wavefield evolution. For receivers near the source, where the direct path is essentially free of scatterers, the term $nT$ naturally equals zero.

Figure 3 illustrates the workflow. For a given transmitter \(N\), we first generate the synthetic field with the forward solver and compute \(\mathrm{d}\!\bigl(\Delta t_{N}^{(k+1)}\bigr)\) for every receiver via \eqref{eq:dt_increment}. Second, starting with the receivers adjacent to the source, we unwrap traveltimes by solving~\eqref{eq:optimize_n}; the corrections run in parallel on both sides of the array and expand outward, as shown in Fig.~3 (left).


The resulting \(\Delta t\) remains reliable at even higher frequencies so long as model updates are modest. Because cumulative error is negligible in the absence of strong multiple scattering, we restrict GRA-TSK to the early, low-frequency stage where its impact is greatest.

\begin{figure}[H]
\vspace{-2mm}
\centering
\includegraphics[width=0.85\linewidth]{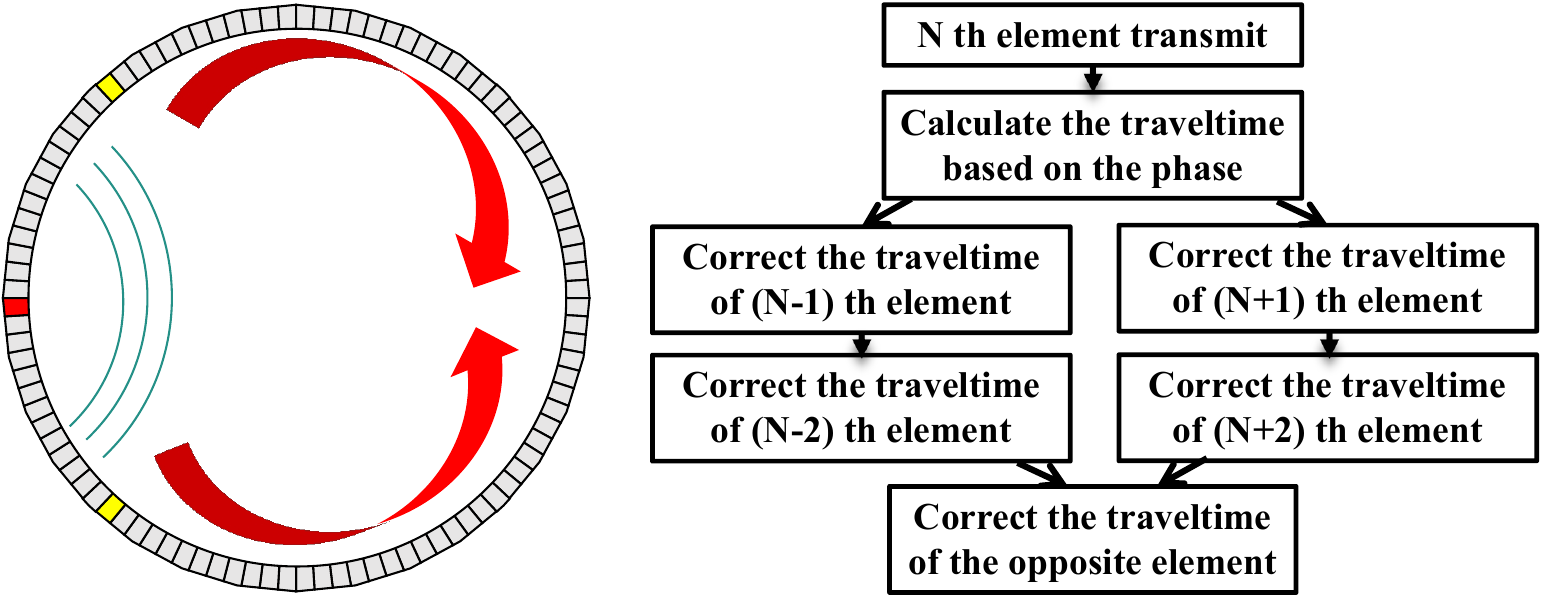}
\label{fig_3}
\vspace{-2mm}
\caption{Parallel correction scheme for traveltime update using adjacent receivers. If $N-i<1$ and $N+i>512$, where $i$ is a positive integer, it refers to the transducer of indexes $N-i+512$ and $N+i-512$, respectively.}
\end{figure}

\subsection{Forward Solver for Helmholtz Equation}
During the inversion process, a forward solver is required to compute the synthetic pressure fields and corresponding adjoint fields for model updates. In FDFWI, one of the most frequently employed solvers is the frequency-domain finite-difference (FDFD) method \cite{ref70}. Other studies have used the convergent Born series (CBS) as the forward solver \cite{ref68}. The HFWI proposed here can be implemented with either approach. In this study, we ultimately adopt the CBS solver.

CBS is a modified approach derived from the Born series, incorporating a relaxation coefficient that is determined by the scattering potential to ensure convergence when applied to arbitrary scattering fields. In contrast to FDFD, whose accuracy relies on the pseudo-inverse of the governing matrix, the accuracy of CBS depends on the number of iterations, providing greater flexibility in practice. Moreover, recent study has demonstrated that CBS can maintain high numerical accuracy even at relatively low points-per-wavelength, reducing the risk of numerical dispersion~\cite{ref_revision_4}. This property allows for coarser discretization without compromising inversion stability, thereby improving overall computational efficiency. Specifically, this work utilizes an open-source CBS package \href{https://github.com/ivovellekoop/wavesim}{(GitHub -- IvoVellekoop/wavesim: High-accuracy simulation of light propagation)}.

\subsection{Inversion Strategy}
\subsubsection{Simulation Setup}

For all numerical and experimental inversions, synthetic and adjoint pressure fields are computed on a $601 \times 601$ grid that spans $0.24\,\mathrm{m} \times 0.24\,\mathrm{m}$. Perfectly matched layers (PML) enclose the domain, and the ring-array element coordinates in Fig.~1(a) are passed directly to the forward solver.


In the numerical study, the observed data are numerically simulated with a time-domain finite difference (TDFD) method~\cite{ref71} on the same physical region and with identical PML settings as the inversion. Each source emits a Ricker wavelet centred at $0.7\,\mathrm{MHz}$ and band-limited to $0.35\text{--}0.9\,\mathrm{MHz}$ by a Butterworth filter (Fig.~4). The resulting grid spacing is about one quarter of the shortest wavelength, ensuring low numerical dispersion, and the time step is set to $30\,\mathrm{ns}$ to satisfy the Courant condition.
\begin{figure}[H]
\vspace{-2.5mm}
\centering
\includegraphics[width=0.95\linewidth]{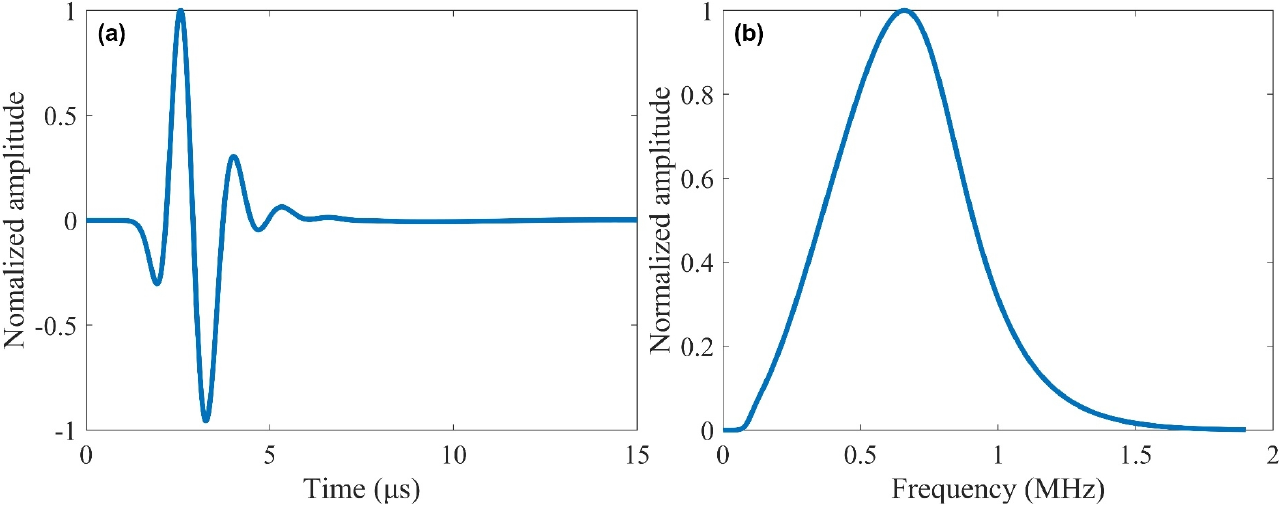}
\vspace{-3.5mm}
\caption{(a) Simulated input pulse used for acoustic excitation. (b) Corresponding frequency spectrum.}
\label{fig_4}
\end{figure}

\subsubsection{Source Estimation}
The phase and amplitude of the simulated source often differ from those implicit in the measured data, which can bias the inversion. We therefore estimate a complex scaling factor for each transmission~\cite{ref41},
\begin{equation}
\gamma = 
\frac{d_{\mathrm{syn}}^{\top} d}
     {d_{\mathrm{syn}}^{\top} d_{\mathrm{syn}}},
\label{eq:src_estimation}
\end{equation}
where $d_{\mathrm{syn}}$ is the synthetic signal $\delta u_{\mathrm{syn}}$—one value per receiver for that shot, as defined in~\eqref{eq:wave_equation}—and $d$ is the corresponding measurement. The calibrated source becomes $S_{\mathrm{syn}}=\gamma\,S_{\mathrm{sim}}$, with $S_{\mathrm{sim}}$ the baseline simulation source.  

The factor $\gamma$ is re-estimated after every model update to track the evolving wavefield. This mean-squared-error matching strategy has proven effective in medical ultrasound tomography~\cite{ref13,ref41,ref55} and is adopted throughout this work.

\subsubsection{Selection of Data}
In experimental tests, trans-target signals are typically weaker because of strong reflections from bone phantoms or limb bones. To enhance the SNR of transmitted waves (i.e., forward scatterings), which are more critical for inversion, the driving voltage of the array is set to $180\,\mathrm{Vpp}$. This may cause backscattered signals to exceed the system’s dynamic range. Additionally, circuit noise introduces crosstalk, such that when one transducer fires, nearby transducers pick up interfering signals, as illustrated in Fig.~5(a), which displays waveforms received by 256 neighboring channels when transducer \#128 is transmitting. Because these noises are difficult to handle in corresponding frequency spectrum as Fig.~5(b), channels from $100$ transducers near the source are discarded during inversion. In addition, early muting is applied to the raw data to suppress system noise before the first-arrival signal. See Supplementary Material Section I for a detailed discussion of the impact of these processing steps.
\begin{figure}[H]
\vspace{-2mm}
\centering
\includegraphics[width=\linewidth]{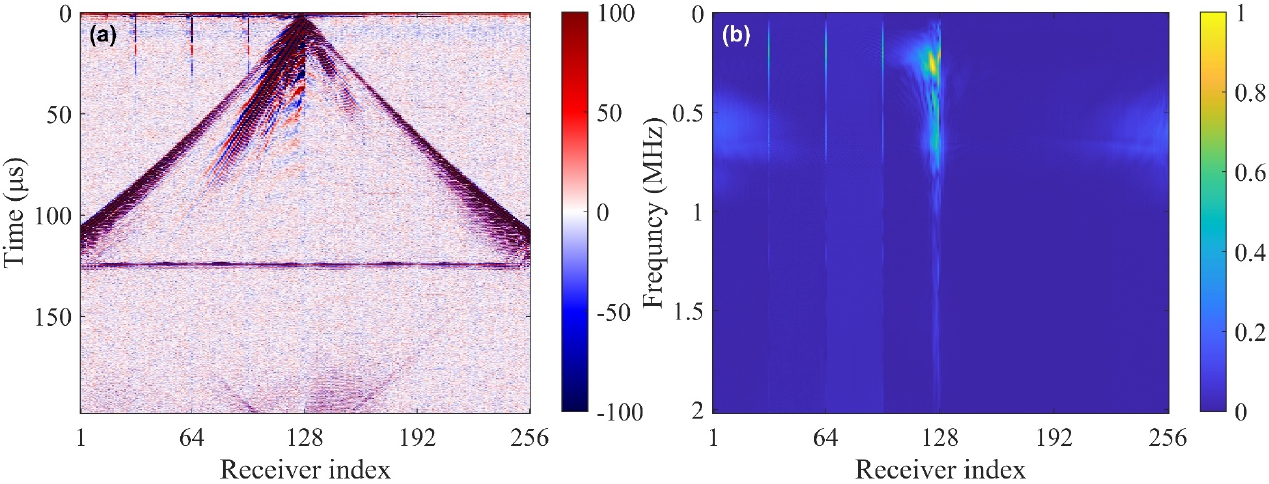}
\label{fig_5}
\vspace{-7mm}
\caption{(a) Experimental FMC data (Pa), showing system noise and crosstalk near the active transmitter. (b) Corresponding normalized frequency spectrum.}
\end{figure}

\subsubsection{Selection of Frequency Points for Inversion}
The overall concept of HFWI parallels that of FDFWI, employing a multiscale inversion strategy that progresses from low to high frequencies. Guided by the actual bandwidth of the ring array, we use frequencies ranging from $0.25\,\mathrm{MHz}$ to $1.2\,\mathrm{MHz}$. To leverage low-frequency signals for mitigating cycle skipping issues while controlling computational costs, frequency points are sampled at $0.05\,\mathrm{MHz}$ intervals between $0.25\,\mathrm{MHz}$ and $0.45\,\mathrm{MHz}$, and at $0.1\,\mathrm{MHz}$ intervals from $0.5\,\mathrm{MHz}$ to $1.2\,\mathrm{MHz}$. The inversion result from each lower-frequency band serves as the initial model for the subsequent higher-frequency inversion, thereby implementing a multiscale strategy.

\subsubsection{Hyperparameters for Hybrid Inversion}
As the inversion proceeds from low to high frequency points, the reconstructed model gradually develops stronger contrasts, particularly in regions corresponding to high-speed structures such as cortical bone. At this stage, the linear relationship between phase shifts and first-arrival traveltimes, as described by equation~\eqref{eq:delta_t}, may no longer hold due to the increasing influence of strong scatterers.

To ensure that the traveltime-based guidance remains reliable, we assign a nonzero value to the hyperparameter $\alpha$ only during the early, low-frequency stages of inversion. This enables GRA-TSK to effectively influence the inversion process while the model remains relatively smooth and the linear approximation remains valid.

Specifically, $\alpha$ is set to $0.75$ at $0.25\,\mathrm{MHz}$ and is progressively reduced with increasing frequency. This gradual transition shifts the inversion scheme from the hybrid formulation back to standard FDFWI. By adopting this strategy, we retain the robustness of GRA-TSK at low frequencies while fully leveraging the high-resolution capability of FDFWI at mid-to-high frequencies, where the SNR is decent.


\subsection{Computational Hardware}
All computational tasks in this study are performed on a workstation equipped with dual Intel Xeon 5812 CPUs and an NVIDIA RTX A6000 GPU (48\,GB). The A6000 provides up to 38.7\,TFLOPS of single-precision (FP32) performance.

\section{Results}

\subsection{Numerical test~I – canonical phantom case}

We first evaluate conventional FDFWI, GRA-based traveltime inversion (GRA-TI), and the proposed HFWI on a canonical phantom at five early-stage frequencies, $\left[0.25,\,0.3,\,0.35,\,0.4,\,0.45\right]$~MHz. The phantom contains three concentric rings with outer diameters of 8, 4, and 2~cm and sound speeds of 1600, 1700, and 2700~m/s, respectively. Its boundaries are smoothed by Gaussian blurring ($\sigma=15$ with a grid spacing of 0.2~mm), and the background is set to 1500~m/s to represent water in a USCT setting, as shown in Fig.~6(a).

The observed data are generated using an FDTD scheme. Because the simulated signals have higher low-frequency SNR than the experimental data, the HFWI weighting parameter is set to $\alpha=\left[0.75,\,0.25,\,0,\,0,\,0\right]$ across the five frequencies, allowing a stronger FDFWI contribution at later stages. This setting is used in all subsequent numerical tests, while pure FDFWI and GRA-TI correspond to $\alpha=0$ and $\alpha=1$, respectively. All inversions are initialized from a homogeneous 1500~m/s model and optimized using nonlinear conjugate gradient, with at most three iterations per frequency and five line searches per iteration. To compensate for the scale mismatch between waveform and traveltime residuals in Eq.~(17), an implementation-level normalization is applied at the beginning of each frequency optimization: the two misfit contributions are initialized as $(1-\alpha)$ and $\alpha$, respectively, while their gradient contributions are separately balanced following Eq.~(19). This numerical scaling improves optimization stability without changing the intended hybrid design of HFWI.


\begin{figure}[t]
    \vspace{-2mm}
    \centering
    \includegraphics[width=0.8\linewidth]{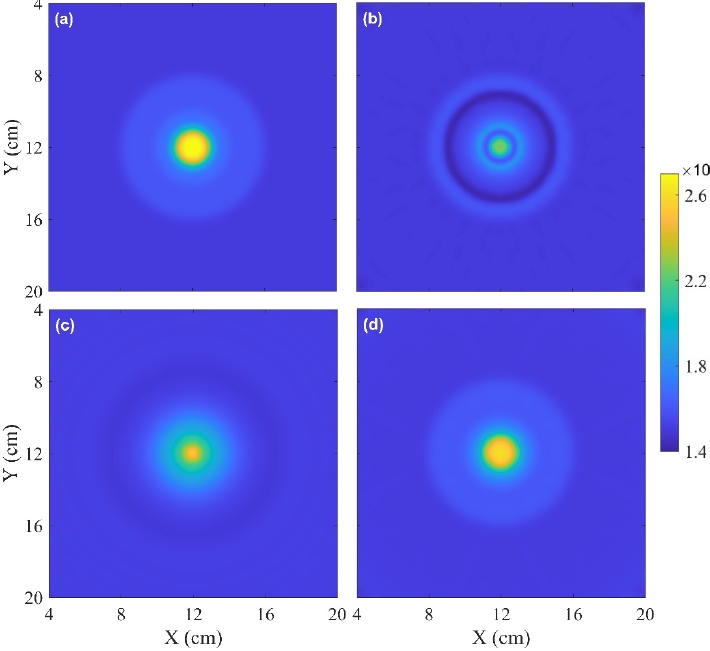}
    \vspace{-4mm}
    \caption{Reconstructed sound speed maps for (b) FDFWI, (c) GRA-TI, and (d) HFWI using the canonical phantom shown in (a). Units: m/s.}
    \label{fig:fig6}
    \vspace{-3mm}
\end{figure}
The reconstructions are shown in Fig.~6(b)--(d), with the corresponding profile at $y=12$~cm shown in Fig.~7. Since the maximum traveltime difference between the initial and true models exceeds 8.8~$\mu$s, more than twice the period at 0.25~MHz, FDFWI is strongly affected by cycle skipping, failing to recover the central high-speed inclusion and producing low-speed artifacts. GRA-TI mitigates this issue by exploiting traveltime information, but its single-frequency implementation provides limited reconstruction accuracy compared with broadband or multi-frequency variants \cite{ref62}. In contrast, HFWI recovers the overall sound speed structure by combining the stability of GRA-TI with the accuracy of FDFWI.

\begin{figure}[h]
    \vspace{-3mm}
    \centering
    \includegraphics[width=0.60\columnwidth]{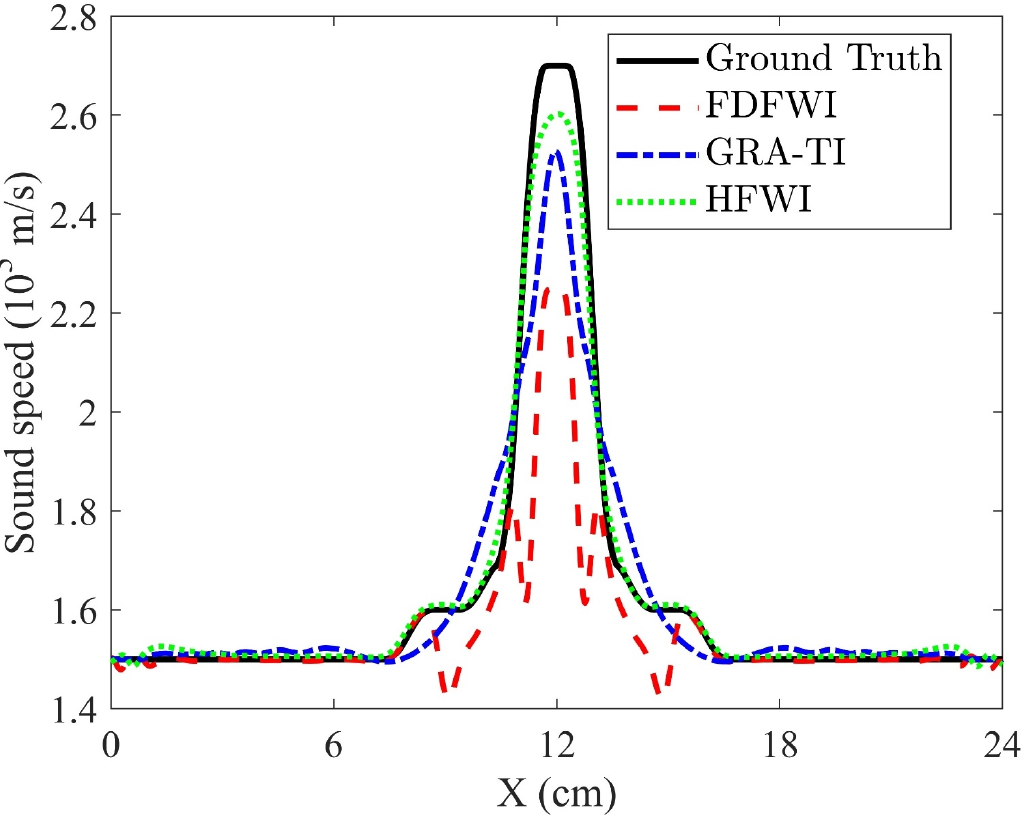}
    \vspace{-3mm}
    \caption{Sound speed profile at $y = 12$~cm for the canonical phantom: ground truth, FDFWI, GRA-TI, and HFWI reconstructions.}
    \label{fig:fig7}
    \vspace{-3mm}
\end{figure}

\subsection{Numerical test~II – anatomical phantom case}
The anatomical phantoms originate from female lower-limb CT scans in the Visible Human Project~\cite{ref72} (Fig.~8), focusing on two-dimensional slices of the left knee and right lower leg. To better approximate real tissue response and account for the acoustic impedance contrast of bone, both sound speed and density are incorporated when generating the observed wavefields. Their acoustic parameters are generated by mapping CT Hounsfield units ($H$) to sound speed and density via the following equation~\cite{ref73}:

The anatomical phantoms are derived from female lower-limb CT scans in the Visible Human Project~\cite{ref72}, using two-dimensional slices of the left knee and right lower leg (Fig.~8). To approximate tissue-dependent acoustic responses and bone-related impedance contrast, both sound speed and density are used when generating the observed wavefields. CT Hounsfield units ($H$) are mapped to acoustic parameters by~\cite{ref73}
\begin{equation}
\mu_x = \mu_{\text{lower}} + \frac{H}{1000} (\mu_{\text{upper}} - \mu_{\text{lower}}),
\label{eq:mapping}
\end{equation}
where $\mu$ denotes either sound speed or density. Although inversion is performed in slowness, all results are displayed in sound speed for consistency with ultrasound imaging studies.

\begin{figure}[htbp]
\vspace{-4mm}
\centering
\includegraphics[width=0.75\linewidth]{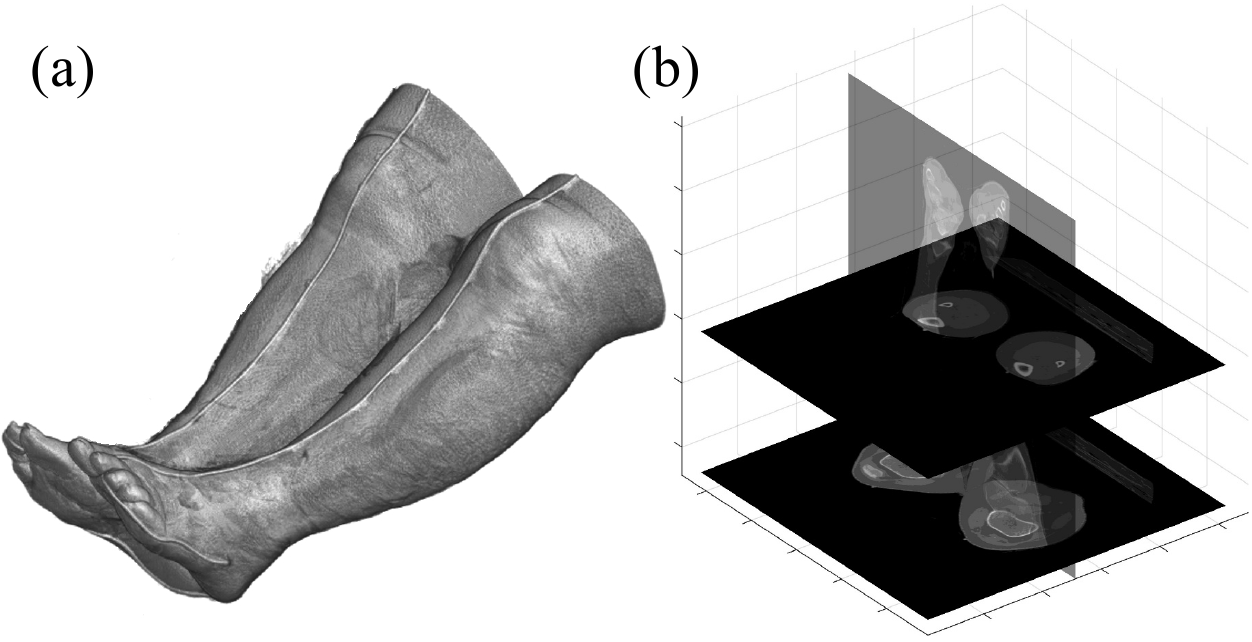}
\label{fig_6}
\vspace{-4mm}
\caption{(a) Volume rendering of the Visible Human Project CT dataset. (b) Selected cross-sectional slices of the knee and lower leg.}
\end{figure}



For the lower-leg section, the sound-speed bounds are set to 1400~m/s for adipose tissue and 3500~m/s for cortical bone, with corresponding densities of 955 and 1900~kg/m$^3$. The resulting sound speed map is shown in Fig.~10(a); the density map is not shown separately because it follows the same CT-derived spatial distribution with a different parameter scale. The observed data are simulated with spatially varying sound speed and density, whereas the inversion assumes a constant density of $1000~\mathrm{kg\,m^{-3}}$ and updates only slowness ($s=1/c$). The resulting amplitude mismatch is largely mitigated by the per-iteration source-estimation step, which rescales the synthetic data to the observations and has been validated in musculoskeletal FDFWI studies~\cite{ref55}. Elastic effects are also neglected, following Ref.~\cite{ref55}, where their influence was shown to be mainly confined to cortical-bone interfaces and limited in soft-tissue regions, in line with the objectives of this study.


Figure~9(a) shows the pressure field recorded by the array for a representative single-element transmission in this lower-leg case, and Fig.~9(b) shows the corresponding first-arrival times extracted for HFWI. The time-domain signals are then transformed to the frequency domain. Both FDFWI and HFWI use the same inversion configuration described in Section~II, with two frequency sweeps from 0.25 to 1.2~MHz; the second sweep starts from a blurred version of the first-round reconstruction. Unless otherwise stated, each frequency uses at most five iterations and five line searches per iteration.

\begin{figure}[htbp]  
    \centering
    \vspace{-4mm}
    \includegraphics[width=\linewidth]{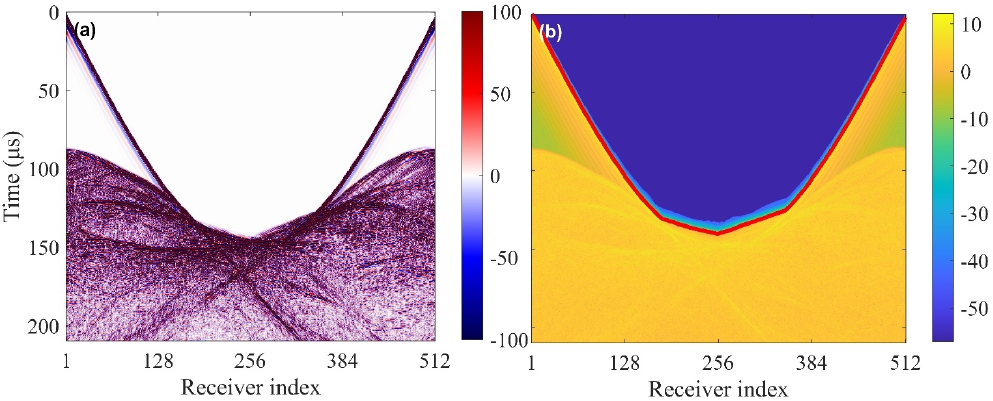}
    \vspace{-7.5mm}
    \caption{Pressure field and first-arrival traveltime picking. (a) Pressure field (Pa) for a representative single-element transmission. (b) Log-scaled waveform data (dB) displayed in dB for visualization, with the extracted first-arrival times overlaid as a red curve.}
    \vspace{-1mm}
    \label{fig:tof_results}
\end{figure}

\begin{figure}[htbp]  
    \centering
    \includegraphics[width=0.75\linewidth]{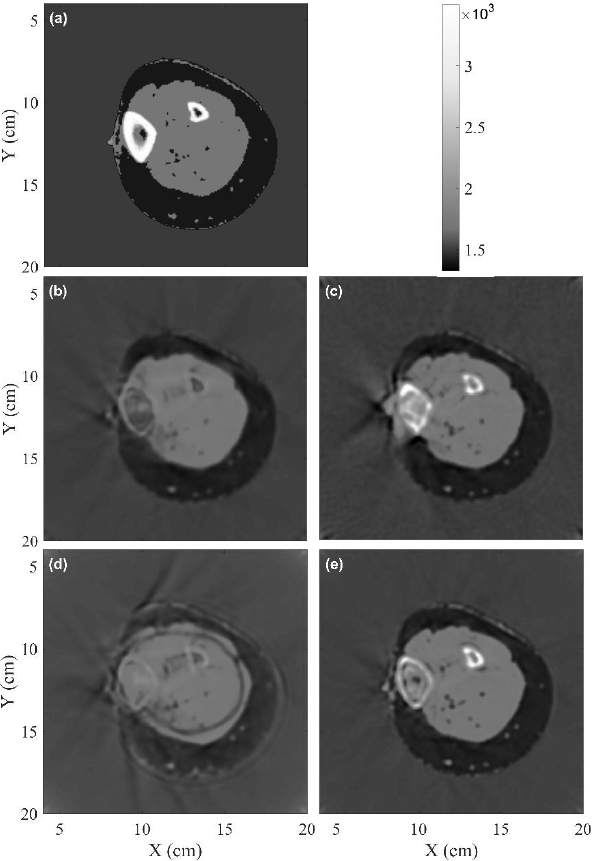}
    \vspace{-4mm}
\caption{Lower-leg numerical phantom case. (a) Ground-truth sound speed from the lower-leg CT slice.
(b), (c) FDFWI and HFWI reconstructions from a homogeneous initial speed of 1500~m/s.
(d), (e) Corresponding reconstructions from an initial speed of 1600~m/s. Units: m/s}
\vspace{-2mm}
\end{figure}

\begin{figure}[!t]  
    \centering
    \includegraphics[width=0.85\linewidth]{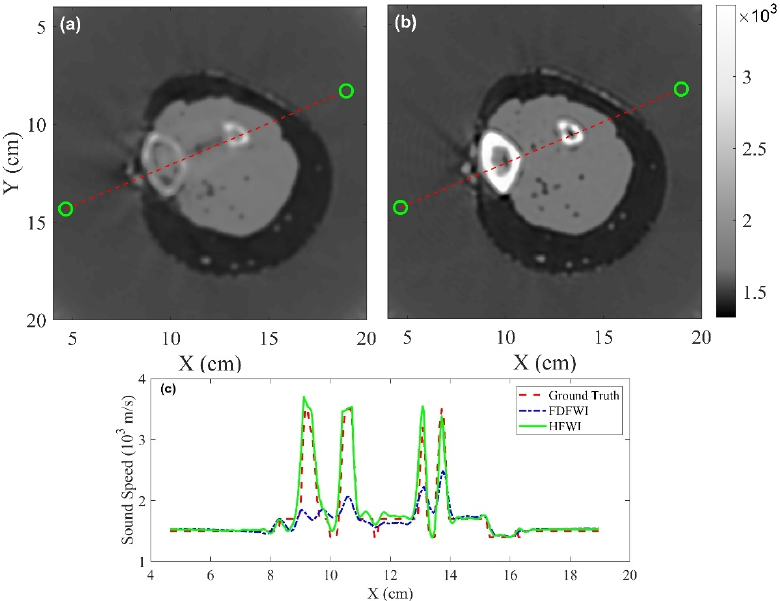}
    \vspace{-4mm}
    \caption{(a) FDFWI reconstruction with extended low-frequency range down to 0.15 MHz.
(b) HFWI result under the same frequency range.
(c) Line profiles of reconstructed sound speed along the green dashed line in (b), highlighting the improved accuracy of HFWI over FDFWI. Units: m/s.}
\end{figure}


Figs.~10(b)--(e) compare FDFWI and HFWI reconstructions from homogeneous initial models of 1500 and 1600~m/s, corresponding to maximum traveltime differences of 14.0 and 9.5~$\mu$s relative to the ground truth. FDFWI fails to resolve the tibia and fibula boundaries and introduces pronounced artifacts. HFWI recovers the bone contours more reliably, including the inner and outer cortical surfaces of the tibia, although residual errors remain near cortical interfaces. This limitation is mainly attributed to the restricted bandwidth, which causes the reconstruction to emphasize scattered wavefields near bone surfaces and leads to edge-enhanced structures.



To further evaluate early-stage model building, an additional lower-leg test is performed from a 1500~m/s background with extended low-frequency components at 0.15 and 0.2~MHz. Compared with Fig.~10(b), FDFWI is improved but still fails to recover the cortical interior accurately (Fig.~11(a)). In contrast, HFWI produces a reconstruction that closely matches the ground truth, as confirmed by the profile comparison in Fig.~11(c).

For the knee section, a slice near the joint is selected. This case contains a thick adipose layer and only thin cortical-bone regions, leading to a limited spatial extent of high-speed structures. The soft-tissue acoustic parameters are adjusted to maintain a challenging musculoskeletal contrast pattern. The ground-truth map in Fig.~12(a) shows irregular bone morphology and fine muscle structures. Starting from a homogeneous 1500~m/s model, the maximum traveltime difference relative to the ground truth reaches 16.2~$\mu$s. As shown in Figs.~12(b) and 12(c), FDFWI mainly produces a narrow peripheral rim, whereas HFWI better captures the anatomical structures. Frequencies below 0.25~MHz are not used in this knee experiment.

Additional quantitative comparisons using RMSE and SSIM are provided in the Supplementary Material as a reference.


\begin{figure}[!t]
\vspace{-3mm}
\centering
\includegraphics[width=\linewidth]{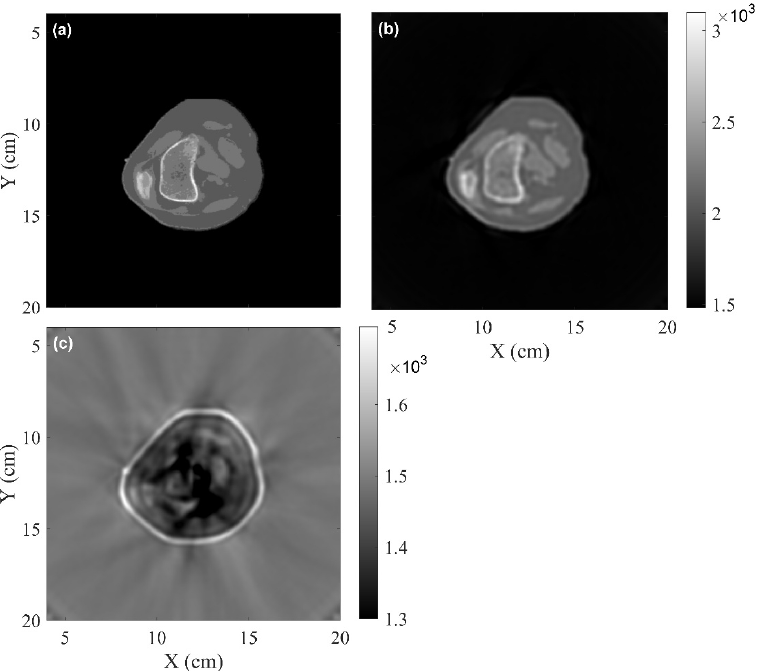}
\label{fig_10}
\vspace{-8mm}
\caption{(a) Sound speed distribution from the selected knee CT slice.
(b), (c) HFWI and FDFWI reconstructions, both initialized with a homogeneous 1500 m/s model. Subfigures (a) and (b) share the same color scale, while (c) is displayed with an independent colorbar for visualization clarity. Units: m/s.}
\vspace{-3mm}
\end{figure}

\subsection{Experimental test – \textit{in vitro} case}

We perform \textit{in vitro} experiments to assess HFWI under limited low-frequency content. As shown in Fig.~13, the phantom consists of a 3D-printed cylindrical resin structure (sound speed $\approx 2800$~m/s) embedded in a soft-tissue-mimicking material (sound speed $\approx 1550$~m/s, with attenuation). The outer boundary is elliptical to avoid geometric symmetry, and the hollow resin structure allows different fluids to be injected. Although the resin speed is lower than that of typical cortical bone and the two resin inclusions do not replicate the tibia--fibula geometry, the phantom provides a controlled experimental case with stronger realism than numerical simulations.

\begin{figure}[!t]
\centering
\includegraphics[width=0.45\linewidth]{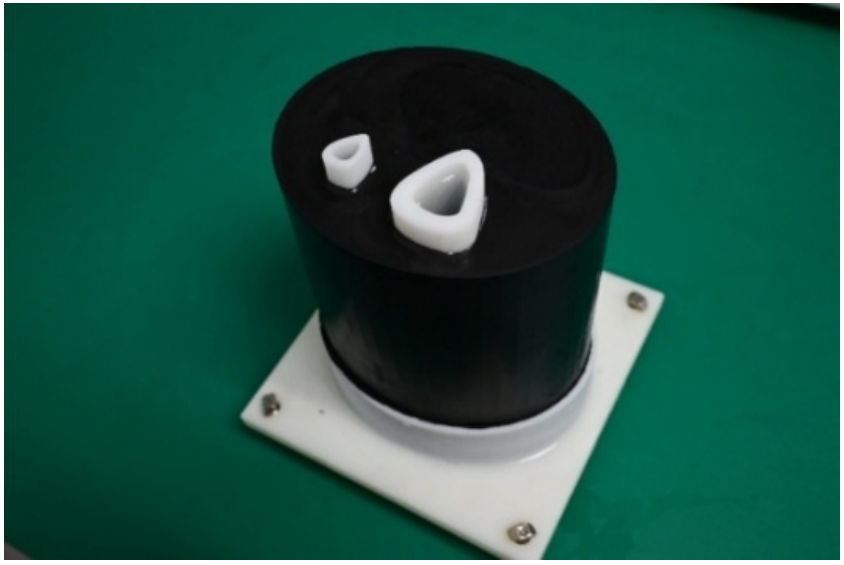}
\vspace{-2mm}
\label{fig_11}
\caption{Photograph of the customized \textit{in vitro} phantom, composed of soft-tissue-mimicking material and embedded bone-like resin structures. Internal cavities enable fluid infusion to modulate acoustic properties.}
\vspace{-3mm}
\end{figure}

\begin{figure}[!b]  
    \vspace{-4mm}
    \centering
    \includegraphics[width=0.85\linewidth]{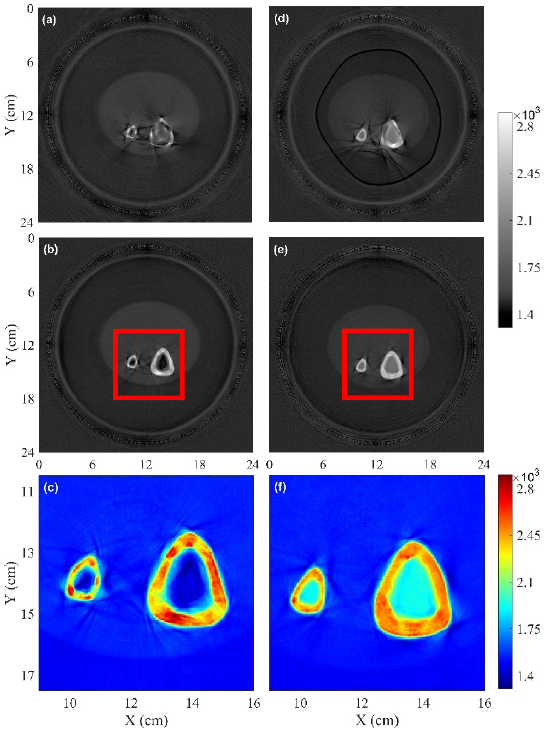}
    \vspace{-4mm}
    \caption{(a), (b) Reconstruction results using FDFWI and HFWI, respectively, with water infused inside the phantom’s bone-mimicking structures.
(c) Magnified view of (b), highlighting the reconstructed bone boundaries.
(d), (e) FDFWI and HFWI reconstructions after replacing the internal fluid with glycerin.
(f) Magnified view of (e), showing improved accuracy in internal sound speed estimation and structural fidelity achieved by HFWI. Units: m/s.}
\end{figure}

\begin{figure}[!t]  
    \centering
    \includegraphics[width=0.85\linewidth]{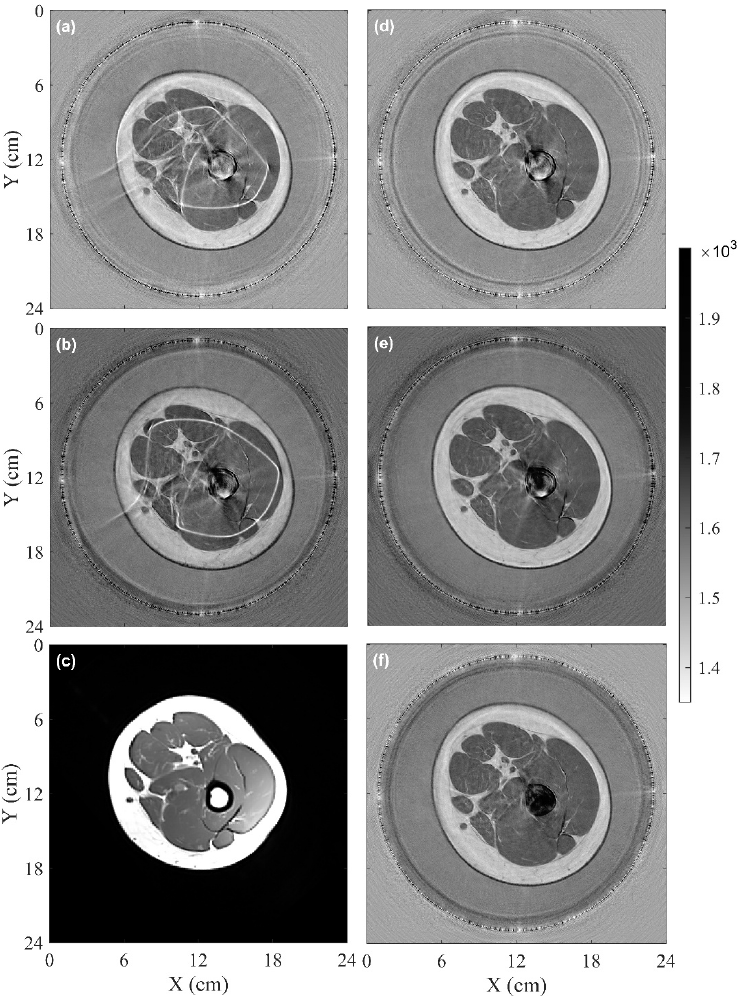}
    \vspace{-3.5mm}

    \caption{(a, d) FDFWI results with an initial speed of 1500 m/s for inversion rounds 1 and 2, respectively. (b, e) FDFWI results with an initial speed of 1600 m/s for inversion rounds 1 and 2, respectively. (c) A 3T T1-weighted MRI slice approximately aligned with the inversion plane, shown for anatomical reference. (f) HFWI result for inversion round 1 with an initial speed of 1500 m/s. Sound-speed maps are in m/s.}
    \vspace{-7mm}
\end{figure}


Compared with the numerical tests, this experiment introduces directional transducer fields, boundary-related propagation effects, and substantially lower SNR, especially in the low-frequency range. In addition, the first-arrival traveltimes extracted by STA/LTA are less accurate than those in simulations. Therefore, the HFWI weighting parameter is adjusted: at 0.25~MHz, $\alpha$ is set to 0.75 for the first three iterations and 0.5 for the last two; at 0.3~MHz, it is set to 0.25 for the first three iterations and 0 for the last two. For all higher frequencies, $\alpha=0$.


We first fill the resin structure with water. Figure~14(a) shows the two-round FDFWI result from a 1500~m/s initial model, which recovers the soft-tissue region but introduces bone-related artifacts between the resin structures. In contrast, HFWI suppresses these artifacts and provides a more stable reconstruction, as shown in Fig.~14(b). The reduced speed contrast relative to cortical bone also alleviates the boundary overemphasis observed in the numerical lower-leg case.


We then replace water with glycerin (sound speed $\approx 1900$~m/s), further reducing the impedance contrast. As shown in Figs.~14(d) and 14(e), severe artifacts remain in FDFWI, whereas HFWI more accurately reconstructs the geometry and the internal glycerin-filled region. The zoomed-in views in Figs.~14(c) and 14(f) further show that the HFWI reconstructions are more consistent with the known phantom structure.

\subsection{Experimental exploration – \textit{in vivo} case}
To evaluate the proposed approach under realistic acquisition conditions, we use cross-sectional data acquired from the upper thigh of a female candidate. Data are collected in a water-tank setup similar to that shown in Fig.~1(b), with the subject standing upright while the ring array is mechanically positioned around the thigh. Under water-immersion conditions, FMC data are acquired for a cross-sectional slice in a manner similar to that illustrated in Fig.~8(b).


We first perform two FDFWI reconstructions using homogeneous initial sound speeds of 1500~m/s, approximating water at 30~$^\circ$C, and 1600~m/s, close to muscle. Both cases use the two-round inversion strategy. The first-round results in Figs.~15(a) and 15(b) exhibit ring-like low-speed artifacts near the muscle--adipose boundary. Following the standard protocol, these outputs are blurred before the second round. As shown in Figs.~15(d) and 15(e), the second round reduces these artifacts and improves the soft-tissue reconstruction, although the femoral region remains difficult to resolve. A 3T T1-weighted MRI slice approximately aligned with the inversion plane is provided in Fig.~15(c) as an anatomical reference.


Next, we perform a single-round HFWI reconstruction using an initial sound speed of 1500~m/s and the same frequency set. This reduces soft-tissue artifacts and yields a clearer reconstruction, achieving soft-tissue imaging quality comparable to the two-round FDFWI results in Figs.~15(d) and 15(e). However, the femoral structure remains only weakly resolved, likely because strong attenuation by the femur weakens the transmitted signals and limits the data constraint in this region under the current acquisition conditions. Runtime information and computational breakdown for this experiment are provided in the Supplementary Material.

\section{Discussion}
Low-frequency content is critical for stabilizing FWI and mitigating cycle skipping, particularly in anatomically complex musculoskeletal regions. However, sufficiently low-frequency signals are difficult to obtain in practical USCT systems without hardware-level improvements. This study therefore focuses on limited-bandwidth conditions and evaluates HFWI as a more stable early-stage model-building strategy than conventional FDFWI.

\subsection{Sensitivity of FDFWI to Initial Model Selection}

In USCT-FWI, homogeneous initial models are commonly used for practical implementation. Although accurate spatial priors can guide FDFWI toward a more reliable convergence basin, they usually require auxiliary inversion steps, such as traveltime-tomography iterations, or manual intervention, such as delay-and-sum-assisted bone initialization. In practice, a constant sound speed between $1450$ and $1650$~m/s is often selected according to approximate tissue properties, mainly to reduce source--receiver phase discrepancies and mitigate cycle skipping.


However, this approximation can still fail when the homogeneous initial model produces large phase errors. As shown in Figs.~6(b) and 12(c), FDFWI can be driven toward incorrect updates when the initial sound speed is inconsistent with the actual traveltime distribution. In these cases, phase mismatch accumulates over iterations, leading to nonphysical structures such as low-speed interiors, high-speed boundary rings, or severe artifacts.


Figure~16 further illustrates this sensitivity by plotting the data misfit against different initial sound speeds. The local extrema near $1500$ and $1570$~m/s indicate that small changes in the starting model can lead to opposite update directions and substantially different erroneous reconstructions.

\begin{figure}[H]
\vspace{-4mm}
\centering
\includegraphics[width=0.6\linewidth]{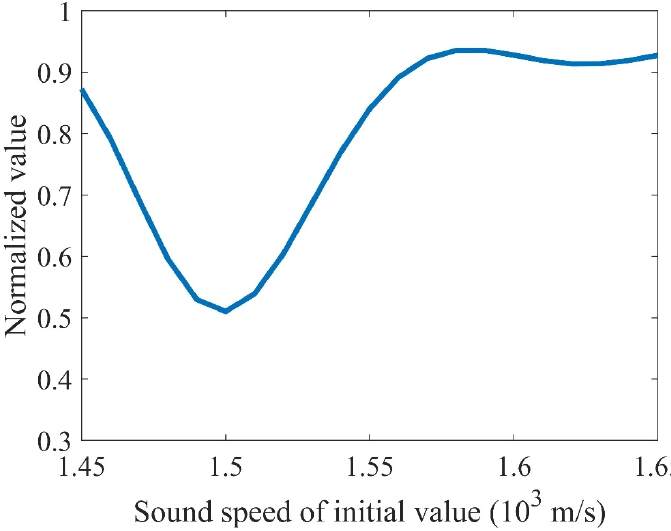}
\vspace{-3mm}
\caption{Misfit values associated with different homogeneous sound speeds.}
\label{fig_14}
\vspace{-3mm}
\end{figure}

\subsection{Limitations of FDFWI in High-Contrast USCT Cases}

FDFWI exhibits two major limitations in high-contrast USCT cases, particularly when the cross section contains multiple high-speed inclusions. The first is an incorrect initial update direction. As shown in Fig.~10(b), both the tibia and fibula are reconstructed with sound speeds lower than the surrounding muscle region, while only a thin outer cortical layer reaches moderately elevated values. This indicates that FDFWI, driven by local waveform matching, can favor small but incorrect updates around the homogeneous initial model.

Figure~17 further illustrates this issue by showing the first-frequency update and final inversion result for FDFWI and HFWI from a homogeneous 1500~m/s initial model. Because the first NCG iteration reduces to a steepest-descent step, the first-update maps in Figs.~17(a) and 17(c), displayed in sound speed after conversion from slowness, visualize the gradient-driven descent directions of FDFWI and HFWI, respectively. In FDFWI, the high-speed regions are driven toward lower sound speeds rather than the expected values, suggesting convergence toward an incorrect local minimum; this behavior persists in the final result in Fig.~17(b). A similar tendency is observed in the \textit{in vivo} case, where the femoral interior remains underestimated in FDFWI (Figs.~15(a) and 15(d)). The corresponding HFWI misfit evolution is shown in Fig.~18.

\begin{figure}[!t]
\vspace{-2mm}
\centering
\includegraphics[width=\linewidth]{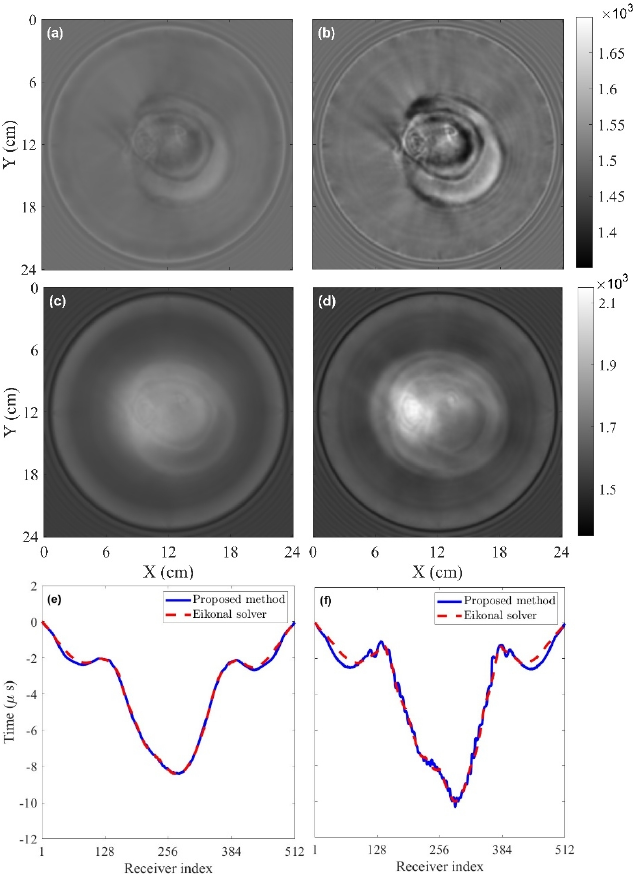}
\label{fig_15}
\vspace{-6mm}
\caption{Comparison of FDFWI and HFWI at 0.25 MHz with the same homogeneous initial model (1500 m/s).
(a) Initial FDFWI update showing incorrect bone speed drop.
(b) Final FDFWI result with artifacts and poor recovery.
(c) Initial HFWI update with correct direction.
(d) Final HFWI result with accurate bone recovery and fewer artifacts.
(e), (f): Traveltime updates relative to the homogeneous background, computed from models (c) and (d), respectively, for source~\#1. Sound-speed maps are in m/s.
}
\vspace{-3mm}
\end{figure}

\begin{figure}[htbp]
\vspace{-2mm}
\centering
\includegraphics[width=0.5\linewidth]{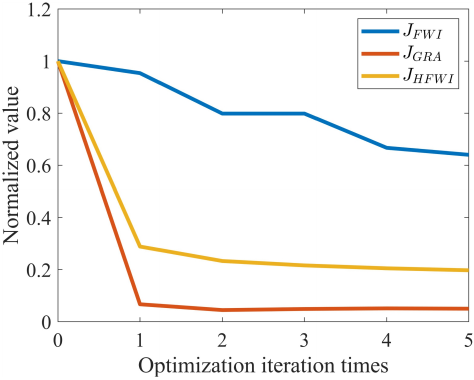}
\vspace{-3mm}
\caption{Data-misfit evolution of HFWI at the first frequency (0.25 MHz), initialized from a homogeneous model of 1500~m/s. Rapid early-stage decrease confirms the effectiveness of the initial update direction and the improved convergence behavior compared to conventional FDFWI.}
\label{fig_18}
\vspace{-3mm}
\end{figure}


The second limitation is the substantial underestimation of bone-related sound speeds even when the initial update direction is improved. As shown in Fig.~10(d), with a more favorable initial model, FDFWI can avoid the most severe wrong-direction updates and recover high-speed regions with values above those of the surrounding muscles. However, the reconstructed values remain far below those expected for cortical bone. Such underestimation may contribute to low-speed artifacts near tissue--bone interfaces and between adjacent high-speed structures. This effect is more evident in the two \textit{in vitro} cases, where FDFWI produces low-speed regions and distorted structures between the two resin inclusions that persist through subsequent iterations, as shown in Figs.~14(a) and 14(d).

\subsection{Role of HFWI in Early-Stage Model Building}

Compared with conventional FDFWI, HFWI improves early-stage inversion robustness by incorporating traveltime information into the waveform inversion process with limited additional computational cost, as indicated by the single-round runtime comparison provided in the Supplementary Material. This effect is first demonstrated in the canonical phantom case (Figs.~6 and 7), where HFWI alleviates the two FDFWI limitations discussed above.

GRA-TSK helps formulate a more linearized update by using traveltime information to reduce large phase discrepancies. As shown in Fig.~18, the first HFWI iteration reduces the global traveltime difference to approximately 10\% of its initial value, suppressing phase errors that would otherwise trigger cycle skipping in the subsequent waveform-driven component. In later iterations, the GRA component acts as a data-driven, projection-like constraint onto a feasible subspace defined by traveltime-consistent updates, rather than as a conventional model-space regularizer, thereby guiding FDFWI along a more stable and less non-convex descent path. As shown in Figs.~17(b)--17(d), HFWI maintains stable updates in high-speed regions and avoids the erroneous FDFWI updates. This also explains why, in the extended low-frequency test (Fig.~11), the additional 0.15--0.2~MHz components help HFWI recover the cortical walls, whereas FDFWI remains affected by incorrect early-stage updates. Figs.~17(e) and 17(f) further compare the traveltime updates relative to the homogeneous initial model obtained using the proposed phase-based update and a conventional Eikonal solver.


This data-driven regularizing effect makes HFWI less sensitive to the homogeneous initial model and facilitates the recovery of high-speed regions, as shown in the lower-leg and knee cases. In highly nonlinear FWI problems, large sound-speed contrasts require multiple iterations to evolve from a homogeneous initial value toward the target distribution. Conventional FDFWI may therefore remain trapped in early-stage mismatches, producing inter-structure scattering errors and low-speed artifacts, as seen in Figs.~10(c), 10(e), 14(a), and 14(d). HFWI mitigates this issue by first introducing a traveltime-guided correction through GRA-TSK, after which the FDFWI component refines the model and improves spatial localization. Consequently, HFWI suppresses inter-structure artifacts in simulations (Figs.~10(d) and 10(f)) and \textit{in vitro} experiments (Figs.~14(b) and 14(e)).

\subsection{Limitations and Practical Considerations of HFWI}

Despite its improved robustness, HFWI retains limitations associated with the assumptions of the GRA-TSK framework. First, the generalized Rytov approximation relates phase shifts to first-arrival traveltime changes in an approximately linear manner, which may become less accurate in the presence of strong scatterers such as bone. Second, the proposed update of $\Delta t$ depends on the initial traveltime differences $\Delta t^{(0)}$ extracted by STA/LTA; errors in this estimate may affect the reconstruction of mid-to-high-frequency features.

These limitations motivate the simultaneous hybrid formulation in HFWI. The frequency-dependent weight $\alpha$ in Eq.~\eqref{eq:HFWI_grad_adj} assigns a larger contribution to GRA-TSK at low frequencies, where the approximation is more reliable, and gradually reduces its influence as the inversion proceeds. This design balances the stability of GRA-TSK with the higher spatial resolution of FDFWI, particularly during early-stage cycle-skipping mitigation.

Another practical limitation is that GRA-TSK mainly captures forward-scattering information. Its updates therefore tend to concentrate near the central propagation region, providing weaker correction near the transducers when the homogeneous initial model deviates from the water sound speed. Although HFWI can still converge after multiple iterations, as shown in Fig.~10(e), convergence is more efficient when the initial model better matches the actual background water speed.

\subsection{Practical Implications for \textit{In Vivo} Reconstruction}
The \textit{in vivo} thigh experiment highlights a practical limitation of conventional FDFWI. Starting from homogeneous initial models, FDFWI tends to generate fluctuating updates near muscle--adipose interfaces, likely due to the sharp contrast between these tissues. As illustrated by the early update patterns in Figs.~17(a) and 17(b), these interface-related errors persist during the first inversion round and evolve into low-speed bands in Figs.~15(a) and 15(b). They are reduced only after a second inversion round, where the blurred first-round reconstruction is reused as the initial model, leading to the improved soft-tissue distributions in Figs.~15(d) and 15(e).

By contrast, HFWI suppresses these interface-related artifacts at an earlier stage and achieves comparable soft-tissue reconstruction quality within a single inversion round (Fig.~15(f)). The reconstructed muscle and adipose distributions are more consistent with the anatomical reference in Fig.~15(c), suggesting potential value for exploratory \textit{in vivo} soft-tissue assessment.

Although HFWI produces a more plausible high-speed region near the femur than FDFWI, the internal bone structure remains insufficiently resolved. This limitation is consistent with the lower-leg simulation in Fig.~11, where cortical-wall recovery improves only when sufficiently reliable lower-frequency components are available. In real acquisitions, complex propagation and strong attenuation inside bone may further weaken the wavefield information associated with the femoral region, which remains difficult to fully address with the present FWI-based formulation.

\section{Conclusion}
We proposed a hybrid full-waveform inversion framework for limited-bandwidth USCT reconstruction. The method combines GRA-based traveltime information with conventional FDFWI in a multi-frequency iterative scheme, aiming to improve early-stage model building when sufficiently low-frequency data are unavailable. By incorporating the modified GRA-TSK into the adjoint-source formulation, the proposed approach updates first-arrival traveltime differences from forward-scattered phase variations without requiring additional wavefield simulations. The per-iteration computational cost therefore remains comparable to that of standard FDFWI.

Simulation and \textit{in vitro} experiments showed that HFWI provides more stable early-stage updates than conventional FDFWI under limited low-frequency conditions, reducing incorrect updates and inter-structure artifacts while improving the recovery of high-speed regions. The \textit{in vivo} thigh experiment further showed soft-tissue reconstruction quality comparable to two-round FDFWI within a single inversion round, suggesting potential value for exploratory \textit{in vivo} soft-tissue assessment.

Internal bone-structure recovery remains challenging under the current bandwidth and acoustic modelling assumptions. Future work will address broader-band acquisition, improved soft--hard tissue interface modelling, three-dimensional implementation, and more robust first-arrival traveltime extraction.

\bibliographystyle{IEEEtran}
\bibliography{references}

\vfill

\end{document}